\documentclass[aps,prb,reprint,superscriptaddress]{revtex4-2}

\pdfoutput=1
\usepackage[utf8]{inputenc}
\usepackage{amsmath,amssymb,amsfonts,graphicx,verbatim,hyperref, mathtools, bbold, float}
\usepackage{physics}
\usepackage{soul, color}
\usepackage{algorithm, algorithmic}
\hypersetup{
	colorlinks=true,
	linkcolor=red,
	citecolor=blue,
	filecolor=blue,
}

\soulregister\ref{7}  % so that \hl and \st can wrap around \ref
\soulregister\cite{7} % so that \hl and \st can wrap around \cite
\newcommand{\hc}{\mathrm{h.c.}}

\DeclareMathOperator*{\argmin}{arg\,min}
\renewcommand*{\eqref}[1]{Eq.~(\ref{#1})}

\newcommand*{\figref}[1]{Fig.~\ref{#1}}

\newcommand*{\secref}[1]{Sec.~\ref{#1}}

\newcommand*{\appref}[1]{Appendix~\ref{#1}}

\begin{document}

\title{Thermalization at Low Temperatures via Weakly-Damped Multi-Site Baths}
\author{Cristian \surname{Zanoci}}
\email{czanoci@mit.edu}
\affiliation{Department of Physics, Massachusetts Institute of Technology, Cambridge, Massachusetts 02139, USA}
\author{Yongchan \surname{Yoo}}
\affiliation{Department of Physics and Joint Quantum Institute,
University of Maryland, College Park, Maryland 20742, USA}
\author{Brian \surname{Swingle}}
\affiliation{Department of Physics, Brandeis University, Waltham, Massachusetts 02453, USA}

\begin{abstract}

 We study the thermalization properties of one-dimensional open quantum systems coupled to baths at their boundary. The baths are driven to their thermal states via Lindblad operators, while the system undergoes Hamiltonian dynamics. We specifically consider multi-site baths and investigate the extent to which the late-time steady state resembles a Gibbs state at some controllable temperature set by the baths. We study three models: a non-interacting fermion model accessible via free-fermion technology, and two interacting models, the XZ model and the chiral clock model, which are accessible via tensor network methods. We show that, by tuning towards the weak coupling and slow relaxation limits, one can engineer low temperatures in the bulk of the system provided the bath size is big enough. We use this capability to study energy transport in the XZ model at lower temperatures than previously reported. Our work paves the way for future studies of interacting open quantum systems at low temperatures.

\end{abstract}

\maketitle

\section{Introduction}
\label{sec:intro}

% open-system dynamics, computational challenges

Open many-body quantum systems are fundamentally different from closed systems and their dynamics displays a wide variety of phenomena not found in equilibrium setups~\cite{landi2022}. For instance, current-carrying non-equilibrium steady states play a major role in deriving the transport properties of the system~\cite{bertini2020finite}. The coupling to an environment can also significantly alter existing properties of the system, such as its phase of matter~\cite{diehl2008,diehl2010,muller2012}. Despite their rich physics, open systems remain vastly understudied, due to both computational and conceptual challenges associated with describing the quantum many-body system and modeling its interaction with the environment. 

Quantum many-body systems are inherently difficult to study due to the exponential growth of their Hilbert space with system size. The computational complexity is further increased for open systems, since we are now dealing with density matrices, instead of wavefunctions. Additional care is necessary to preserve the positivity and hermiticity of the state. We therefore require methods that are specifically tailored to this setup~\cite{weimer2019simulation}. Some of the most popular choices include tensor networks~\cite{verstraete2008,schollwock2011density,orus2014,bridgeman2017,cirac2021}, quantum trajectories~\cite{daley2014}, and neural networks~\cite{hartmann2019,yoshioka2019,nagy2019,vicentini2019,reh2021,luo2022}. In the context of one-dimensional systems, matrix product state methods~\cite{verstraete2004matrix,zwolak2004,weimer2015,cui2015variational,mascarenhas2015matrix,werner2016positive,brenes2020} have proven particularly powerful for describing large-scale open systems. However, even these techniques can suffer from slow convergence to the steady state if the open system dynamics are not properly designed.

% lindblad approach and its limitations

Efficiently modeling the system's interaction with the environment remains a major challenge within the field. One way to approach this problem is by viewing the target system together with the infinitely large environment as a closed system undergoing Hamiltonian evolution. The evolution of the target system is then recovered by tracing out the environment degrees of freedom. However, the resulting master equations are usually non-local in time and involve a complicated memory kernel~\cite{nakajima1958,zwanzig1960,landi2022}. The master equation simplifies significantly within the Born-Markov approximation, leading to the Redfield equation~\cite{redfield19651}, which is unfortunately not guaranteed to preserve the positivity of the density matrix. A further secular approximation is necessary to mitigate this problem, resulting in a global Lindblad master equation~\cite{lindblad1976generators,gorini1976,landi2022}. Nevertheless, identifying the correct global Lindblad operators leading to the desired dynamics can be computationally impractical for large systems, since it usually requires knowledge of the full energy eigenbasis. 

Due to these limitations, the majority of studies on open many-body quantum systems rely on a local Lindblad description of the dynamics. In this approach, the interaction with the environment is modeled by Lindblad jump operators acting locally on the boundary of the system~\cite{prosen2009matrix,znidaric2010,vznidarivc2011transport,vznidarivc2011spin,znidaric2012,znidaric2013a,znidaric2013b,mendoza2013heat,mendoza2015, mendoza2019asymmetry,zanoci2021}, such that the bulk dynamics is still coherent and governed by the system's Hamiltonian. In the thermodynamic limit, the bulk properties should not depend on the details of the boundary driving, given that the system is sufficiently ergodic. However, the local Lindblad equation can sometimes fail to describe the correct steady state, as reported for integrable systems~\cite{znidaric2010,mendoza2015}, in the weak coupling limit~\cite{wichterich2007,levy2014,xu2017,tupkary2022,tupkary2023}, and in the presence of multiple baths~\cite{rivas2010,purkayastha2016}. Moreover, it is also not clear whether the boundary driving can thermalize the system to arbitrarily low temperatures. 

% previous work on thermalization

The issue of thermalization in open quantum systems has been previously studied in Ref.~\cite{reichental2018}, in the context of both interacting and non-interacting fermionic systems with particle number conservation. Their results, based on a perturbative expansion in the limit of zero system-bath coupling, suggest that both models thermalize if the baths are infinitely large and weakly damped. In this paper, we extend these results to strongly interacting spin chains, whose steady states admit a tensor network representation. Furthermore, we show that a system coupled to a thermal bath at its boundary typically reaches a temperature far above the one imposed by the bath. We then pinpoint the conditions under which the system thermalizes to the desired bath temperature. 

% what we do

We consider three different one-dimensional systems in our analysis. The first one is the complex Sachdev-Ye-Kitaev (SYK) model describing fermions with random all-to-all $2$-body interactions~\cite{sachdev1993,parcollet1999,georges2000,georges2001,sachdev2015,Fu2016,davison2017,Bulycheva2017,Gu2020,tikhanovskaya2021a,tikhanovskaya2021b,chowdhury2021sachdev}. This Hamiltonian is quadratic in the fermionic operators and its steady state can be computed exactly~\cite{Prosen2008,prosen2010}. We use this model to test the validity of the previous results and to determine the conditions under which the system can be cooled to low temperatures by an external bath. The second system is described by a spin-1/2 XZ Hamiltonian in a transverse field~\cite{ye2019emergent}. This model is interacting and non-integrable, and we expect its behavior to be fairly representative of one-dimensional gapped systems. Finally, the third model is known as the chiral clock model and can be tuned to a gapless non-integrable quantum critical point~\cite{huse1981simple,ostlund1981incommensurate,huse1982domain,huse1983melting,haldane1983phase,howes1983quantum,au-yang1987commuting,fendley2012parafermionic,ortiz2012dualities,zhuang2015phase,dai2017entanglement,samajdar2018,whitsitt2018}. In the absence of an analytical solution, the steady states for both of these models are computed via tensor network methods~\cite{verstraete2004matrix,zwolak2004,vidal2003efficient,vidal2004efficient,paeckel2019}. The final temperature of the system is extracted using our previously developed thermometry technique~\cite{zanoci2021}. 

Our results show that the system thermalizes even at low temperatures, as long as the bath is extremely large, weakly coupled to the system, and infinitely damped, in agreement with previous findings~\cite{reichental2018,zanoci2016entanglement}. In practice, however, taking these limits is often unnecessary. We find that convergence to within $1\%$ of the target bath temperature can already be achieved with leads that are a fraction of the system size by lowering the system-bath coupling $g$ and the bath driving strength $\gamma$ by an order of magnitude. This observation has important implications for models whose dynamics is not exactly solvable, where we rely on approximate time-evolution methods to find the steady state. 

% more results on diffusivity

As an immediate practical application, we study energy transport by attaching two boundary baths and imposing a temperature gradient across the system. The temperature bias is small, such that the system is only weakly perturbed from equilibrium. This results in a constant energy gradient and current across the chain. By adjusting the average bath temperature, we are able to extract the temperature dependence of the transport coefficients. In particular, we focus on the low-temperature behavior of the diffusivity in the XZ model and are now able to reach regimes that were previously inaccessible with conventional open system setups~\cite{zanoci2021}. We find that the energy diffusivity increases exponentially at low temperatures, in agreement with the predictions of a semi-classical kinetic theory for gapped systems~\cite{damle1998spin,damle2005universal,zanoci2021}. 

% limitations

The ultimate limiting factor in our study is the minimal temperature below which the baths can no longer reliably cool the system on the timescales accessible numerically. We conjecture that this temperature is set by the smallest energy scale in the problem -- either the model's gap or its interaction strength. Other methods beyond the Lindblad master equation might be able to circumvent this limitation. 

% outline
The outline of this paper is as follows. In \secref{sec:setup} we present our open system setup and describe some key properties of the steady states and the relaxation dynamics towards them. In \secref{sec:results} we introduce our models and show the thermalization results for each of them. We also analyze the low-temperature transport properties of the XZ model. Finally, we comment on our findings and discuss possible extensions in \secref{sec:discussion}.

\section{Setup}
\label{sec:setup}

\begin{figure}
\begin{center}
\includegraphics[width=\columnwidth]{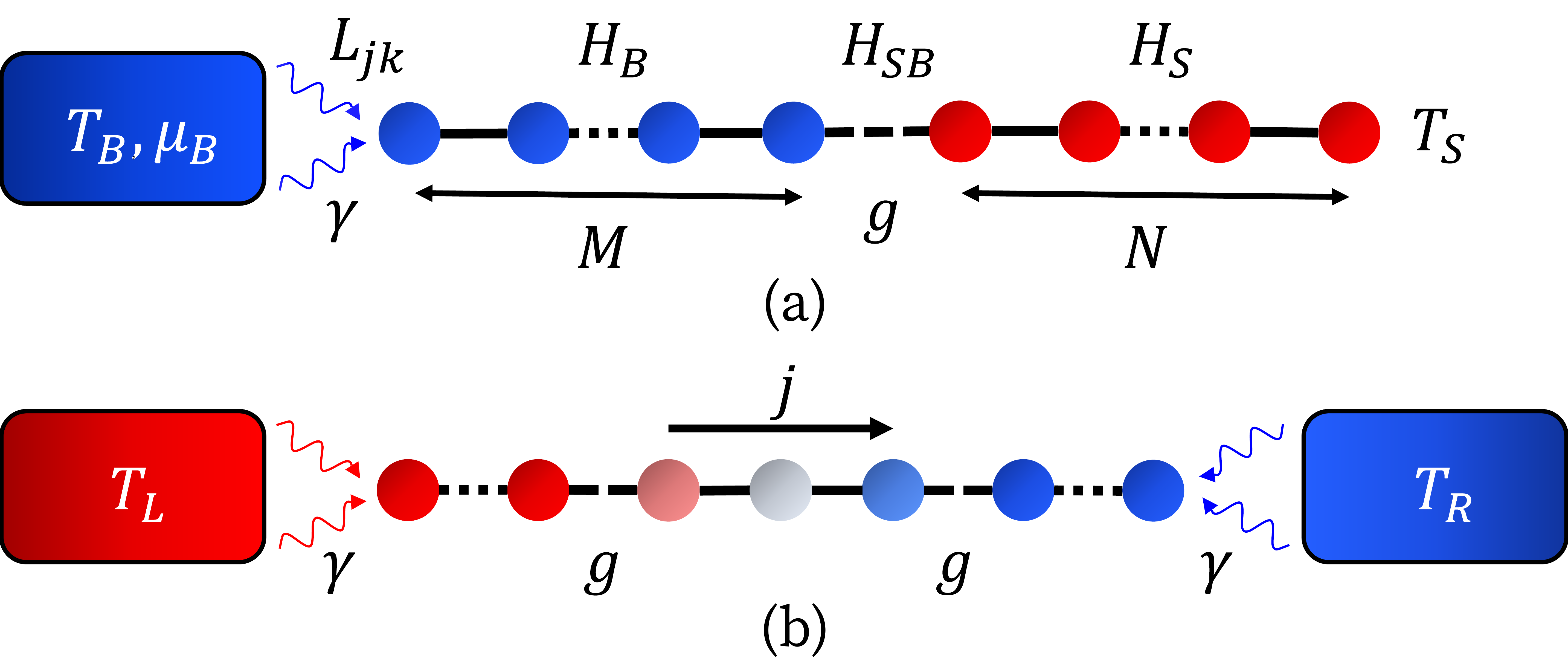}
\caption{Schematic diagram of our (a) thermalization and (b) transport setups. The system is connected to baths at its boundary. The Lindblad operators $L_{jk}$ drive the baths to their thermal states. In non-equilibrium, a homogeneous current $j$ flows through the bulk of the system.}
\label{fig:fig1}
\end{center}
\end{figure}

We begin by presenting the general setting of our thermalization studies, which is depicted in~\figref{fig:fig1}(a). The setup consists of a system $S$ of size $N$ coupled to a bath $B$ of size $M$ at its boundary. The environment only acts upon the bath. The full Hamiltonian is given by

\begin{equation}
    H = H_S + H_B + gH_{SB},
\end{equation}
where $H_S$, $H_B$, and $H_{SB}$ are the Hamiltonians describing the system, bath, and their interaction, respectively, while $g$ is a dimensionless parameter controlling the system-bath coupling strength. The state of the entire open quantum system at time $t$ is characterized by its density matrix $\rho(t)$. The evolution of this density matrix is governed by the local Lindblad equation~\cite{lindblad1976generators,gorini1976}

\begin{equation}
    \frac{d\rho}{dt} = \mathcal{L}(\rho) \equiv -i[H, \rho] + \sum_{jk}\left(L_{jk}\rho L_{jk}^\dagger - \frac{1}{2}\{L_{jk}^\dagger L_{jk}, \rho\}\right), 
    \label{eq:lindblad}
\end{equation}
where $L_{jk}$ are jump operators acting only on the bath degrees of freedom and describing their interaction with the environment. These operators are chosen such that, when decoupled from the system ($g=0$), the bath is driven to its thermal state $\rho_B$ at temperature $T_B$ and chemical potential $\mu_B$. The strength of this damping introduces a new energy scale $\gamma$ into the picture. A generic construction for the Lindblad operators $L_{jk}$ that thermalize baths of any size is described in \appref{sec:appendixA}. 

Initially, the system starts in an arbitrary (usually infinite temperature) state, while the bath is in its thermal state $\rho_B$. Once the coupling $g$ between the two is turned on, the baths are still continuously driven towards their decoupled thermal state, but the coupling with the system causes the steady state of the bath to differ from its target thermal state. The system also begins to exchange energy and particles with the bath while converging to its steady state. The operator $\mathcal{L}$ capturing this time evolution in~\eqref{eq:lindblad} is called the Liouvillian super-operator and it acts on the space of density matrices. For Lindblad equations, this map is completely positive and trace-preserving~\cite{lindblad1976generators,gorini1976}. The steady state $\rho_\infty$ is given by the fixed point of this map $\mathcal{L}\rho_\infty=0$. The relaxation time to this steady state depends on the gap in the Liouvillian spectrum. Assuming that the fixed point of the Lindblad equation is unique, the gap $\Delta_{\mathcal{L}}$ is equal to the negative real part of the second largest eigenvalue of $\mathcal{L}$~\cite{Prosen2008,prosen2009matrix,znidaric2015}.

Formally, the solution of the Lindblad equation is given by the right eigenvector of $\mathcal{L}$ with eigenvalue zero. While closed-form exact solutions exist in the case of non-interacting systems~\cite{Prosen2008,prosen2010} and for certain strongly-driven interacting systems~\cite{clark2010exact,karevski2013exact,prosen2011exact,prosen2011,prosen2014exact,popkov2015,popkov2016,popkov2020exact,popkov2020}, one usually has to resort to approximate methods when dealing with generic interacting systems and arbitrary driving. This is because the dimensions of the Liouvillian grow exponentially with system size. In practice, it is more feasible to directly simulate the time evolution and look for a converged final state $\rho_\infty = \lim\limits_{t\to\infty}\rho(t) = \lim\limits_{t\to\infty}e^{\mathcal{L}t}\rho(0)$. To achieve this, we employ a tensor network representation of the density matrix~\cite{verstraete2004matrix,zwolak2004} and perform an approximate time evolution under the Liouvillian super-operator using the Time Evolving Block Decimation (TEBD) algorithm~\cite{vidal2003efficient,vidal2004efficient,paeckel2019}. We represent the states and operators in vectorized form~\cite{zwolak2004,weimer2019simulation,landi2022} and use a second-order Suzuki-Trotter decomposition~\cite{suzuki1990fractal} of the time-evolution operator $e^{\mathcal{L}t}$. We should mention that alternative approaches, such as variational algorithms targeting the ground state of the Hermitian operator $\mathcal{L}^\dagger\mathcal{L}$~\cite{weimer2015,cui2015variational,mascarenhas2015matrix}, could also be used to find steady state $\rho_\infty$ more directly.

The limiting factor in reaching the steady state is the size of the spectral gap $\Delta_{\mathcal{L}}$. If the gap is finite, the distance between the time-evolved state $\rho(t)$ and the steady state $\rho_\infty$ will decrease exponentially in time, with a relaxation rate set by $\Delta_{\mathcal{L}}$. For systems that are only subject to boundary dissipation, a generic bound on the gap as a function of system size $\Delta_{\mathcal{L}}\lesssim 1/N$ is known~\cite{znidaric2015}. In integrable systems, one typically observes a scaling $\Delta_{\mathcal{L}} \sim 1/N^3$, as is the case for nearest-neighbor hopping free-fermion models~\cite{znidaric2015,zanoci2016entanglement}. On the other hand, for non-integrable models, one usually has a faster relaxation rate which saturates the bound $\Delta_{\mathcal{L}}\sim 1/N$. The gap can also be exponentially small in localized systems~\cite{znidaric2015,zanoci2016entanglement}. Furthermore, perturbation theory calculations in the limit of small system-bath coupling ($g\to0$) or small driving ($\gamma\to0$) reveal the following scaling~\cite{cai2013,znidaric2015,medvedyeva2016} 

\begin{equation}
    \Delta_{\mathcal{L}}\sim\gamma g^2, \quad \text{as} \quad g,\gamma\to0.
    \label{eq:gap}
\end{equation}
In practice, this implies that we cannot use traditional time-evolution methods to study the regime where these parameters are infinitesimally small, as the convergence to steady state would be prohibitively slow. It turns out that this limit is precisely the one required to observe thermalization~\cite{reichental2018}. Nevertheless, we will show in~\secref{sec:interacting} that even moderate values of $g$ and $\gamma$ accessible numerically for small systems are sufficient to reach states that are close to thermal. 

Once we find the steady state, we have to evaluate how well the system resembles a Gibbs state at some temperature and whether this temperature is close to the bath's driving temperature $T_B$. The state of the system alone is obtained by tracing out the bath degrees of freedom $\rho_S=\Tr_B(\rho_\infty)$. Assigning a global temperature to this state requires access to reference Gibbs states of the same size, which is unfeasible for large systems. Moreover, we expect the system to deviate from thermal equilibrium near the boundary where it is coupled to the bath. Therefore, it is more suitable to assign a local temperature for different parts of the system~\cite{hartmann2004,hartmann2005,hartmann2006minimal,garcia2009,ferraro2012intensive,kliesch2014locality}. To accomplish this, we use the thermometry method introduced in Ref.~\cite{zanoci2021}. The local temperature of a region $A$ is derived by minimizing the trace distance between the reduced density matrices of that region in the steady state $\rho_S^A$ and in a thermal state $\rho^A(T)$

\begin{equation}
   D\left(\rho_S^A, \rho^A(T)\right) = \frac{1}{2}\Tr\left(\sqrt{\left(\rho_S^A-\rho^A(T)\right)^2} \right).
   \label{eq:trace_dist}
\end{equation}
The trace distance between the two states is a meaningful metric because it places an upper bound on the difference between the corresponding expectation values of any local observable~\cite{mendoza2015}. We usually choose the region $A$ to be comprised of two consecutive sites $(i, i+1)$ in the system, which is consistent with our definition of local energy. For the models we studied, we found good thermalization and a uniform local temperature away from the boundaries. Therefore, we define the temperature of the system $T_S$ in a steady state as the local temperature at its center

\begin{equation}
   T_S = \argmin_{T} D\left(\rho_S^{\left(\frac{N}{2}, \frac{N}{2}+1\right)}, \rho^{\left(\frac{N}{2}, \frac{N}{2}+1\right)}(T)\right).
   \label{eq:system_temp}
\end{equation}

\section{Results}
\label{sec:results}

The framework described above can in principle be applied to any system. In this work, we consider three prototypical examples of non-interacting and interacting systems in one dimension. We start with a non-interacting SYK2 model, since it has a simple description in terms of single-particle eigenstates and its steady state is exactly solvable even for large system sizes. We use this model to find the precise limits in which the bath thermalizes the system and verify that they agree with previous predictions~\cite{reichental2018}. We then extend our analysis to interacting XZ and chiral clock chains, which do not have analytical solutions, and find that the same thermalization behavior persists for these systems. Additionally, we show how to apply the lessons learned from the equilibrium setup to study low-temperature energy transport in the XZ model. Our numerical results focus on the intermediate ($T_B = 1$) and low ($T_B = 0.1$) temperatures, since we expect thermalization to be especially challenging in these regimes~\cite{zanoci2021}. However, we checked that our conclusions hold for a wide range of temperatures, including the high-temperature limit. Lastly, we set $\mu_B = 0$ for the remainder of this paper.

\begin{figure*}[tp]
\begin{center}
\includegraphics[width = \textwidth]{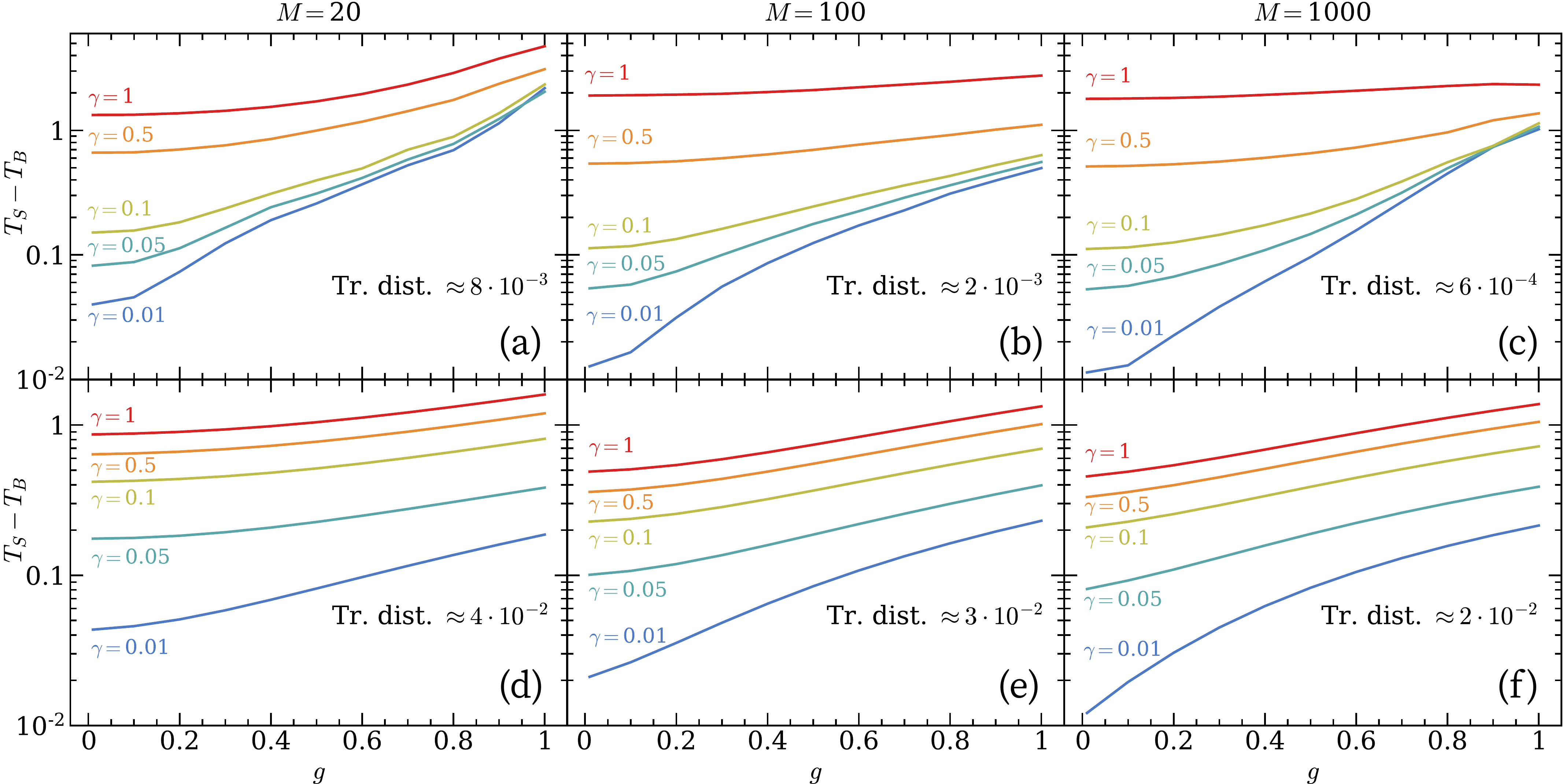}
\caption{Thermalization results for the SYK2 model coupled to a boundary bath at temperatures (a-c) $T_B = 1$ and (d-f) $T_B = 0.1$. The system temperature $T_S$ approaches the bath temperature $T_B$ in the limit of weak system-bath coupling $g$ and damping $\gamma$. The trace distances shown are for the steady states with the lowest temperature.}
\label{fig:fig2}
\end{center}
\end{figure*}

\subsection{Non-interacting SYK2 Model}
\label{sec:non-interacting}

Our first model is a complex SYK2 cluster with random all-to-all $2-$body interactions among spinless fermions~\cite{sachdev1993,parcollet1999,georges2000,georges2001,sachdev2015,Fu2016,davison2017,Bulycheva2017,Gu2020,tikhanovskaya2021a,tikhanovskaya2021b,chowdhury2021sachdev}. Both the bath and the system are described by similar Hamiltonians 

\begin{align}
    \label{eq:SYK_ham}
    H_B &= \sum_{1\leq i,j\leq M}J_{ij}^{B}c_i^\dagger c_j,\\
    H_S &= \sum_{M+1\leq i,j\leq M+N}J_{ij}^{S}c_i^\dagger c_j.
\end{align}
The fermions obey the standard anti-commutation relations $\{c_i^\dagger, c_j\}=\delta_{ij}$. The SYK couplings are complex, independent Gaussian random variables with zero mean obeying 

\begin{align}
    J_{ij}^{B,S} &= (J_{ji}^{B,S})^*, \\
    \langle |J_{ij}^{B}|^2 \rangle = \frac{J_B^2}{M}, & \quad \langle |J_{ij}^{S}|^2 \rangle = \frac{J_S^2}{N},
\end{align}
where the factors of $M$ and $N$ ensure that the energy is extensive. The SYK2 model describes free fermions, since it can be diagonalized in the energy eigenbasis

\begin{equation}
    H_{B,S} = \sum_{k} \epsilon_k^{B,S} c_k^\dagger c_k.
    \label{eq:SYK_ham_energy}
\end{equation}
It is also integrable and non-chaotic~\cite{davison2017}. The coupling between the system and the bath is simply given by 

\begin{equation}
    H_{SB} = J_{SB}(c_M^\dagger c_{M+1} + c_{M+1}^\dagger c_M),
\end{equation}
since we wanted to keep the interaction local and restricted to the boundary of the system. One could also choose an SYK2 interaction between the two sides and recover the same results.

In the non-interacting case, the Lindblad equation is quadratic in the fermionic operators and can be solved analytically using the third quantization technique~\cite{Prosen2008,prosen2010}. This method allows us to study fairly large systems with extremely high accuracy. The main idea is to write the Liouvillian $\mathcal{L}$ in terms of adjoint Majorana maps and diagonalize it in the basis of normal master modes, which represent anticommuting super-operators acting on the Fock space of density operators. The steady state is then given by the zero-mode eigenvector of an antisymmetric matrix of size $4(M+N)$. The reduced density matrix of the system is diagonal in the third quantization eigenbasis

\begin{equation}
    \rho_S = \prod_{k=1}^N \left(n_kc_k^\dagger c_k + (1-n_k)c_kc_k^\dagger\right),
    \label{eq:rho_S}
\end{equation}
where the occupation numbers $n_k$ are sorted in descending order and modes $c_k$ are computed numerically~\cite{zanoci2016entanglement}. In the thermal state of the decoupled system, the occupation numbers are expressed in terms of the Fermi-Dirac distribution

\begin{equation}
    f_k = \frac{1}{1+e^{(\epsilon_k - \mu)/T}}.
    \label{eq:fermi-dirac}
\end{equation}
However, in the limit of weak system-bath coupling, we expect the occupation numbers $n_k$ to be close to the thermal values $f_k$. Therefore, in order to determine the temperature $T_S$ associated with $\rho_S$, we minimize the average trace distance between the single-particle density matrices in the steady state and in thermal equilibrium

\begin{equation}
    D(\rho_S, \rho(T)) = \frac{1}{N}\sum_{k=1}^N |n_k - f_k|.
    \label{eq:trace_dist_2}
\end{equation}
This metric closely matches the one introduced in~\eqref{eq:trace_dist}.

Our results are shown in~\figref{fig:fig2}. We set $J_B = J_S = J_{SB} = 1$, $N = 100$, and average over $100$ realizations of the interaction matrices $J^{B,S}$. We consider three different bath sizes $M = 20, 100, 1000$, which are representative of small, large, and infinite reservoirs, respectively. The absolute deviation of the system's temperature from the target bath temperature is plotted as a function of system-bath coupling $g$ for several rates $\gamma$. We observe that $T_S$ monotonically increases with both of these parameters. At $g = 1$, the system temperature can be up to an order of magnitude larger than $T_B$. In the limit of zero coupling $g$, the temperature deviation converges to a finite value with a substantial $\gamma-$dependence. In order for the system to thermalize at the desired temperature, one must additionally take the limit $\gamma\to0$.  

The error in approximating the steady state solution with a Gibbs state is quantified by~\eqref{eq:trace_dist_2}. Based on this metric (see~\figref{fig:fig2}), we see that achieving good thermalization at low temperatures can be challenging. The trace distance is almost independent of $g$, but does decrease with $\gamma$. Moreover, larger baths lead to steady states that are much closer in trace distance to thermal states. For $M = 1000$, we effectively have an infinite bath, which provides very little improvement over the configuration with $M = 100$. Somewhat surprisingly, the system does not attain perfect thermalization solely in the infinite bath limit. The reason for this becomes more clear in the perturbative expansion discussed below~\cite{reichental2018}. We should emphasize that our findings here do not contradict our previous results showing that SYK2 clusters thermalize when connected to infinitely large reservoirs with no backreaction~\cite{almheiri2019,zanoci2022,zanoci2022_2}, which were derived in quite a different setup not involving Lindblad open system dynamics. 

In order to gain an analytical understanding of the interplay between various parameters in the problem, we seek a perturbative solution for the steady state at small $g$ and $\gamma$~\cite{cai2013,znidaric2015,medvedyeva2016}. As argued in Ref.~\cite{reichental2018}, it is more natural to take the limit of $g\to0$ first. Following their derivation, the system's reduced density matrix is of the form introduced in~\eqref{eq:rho_S} with occupations

\begin{equation}
    n_k = \frac{\sum_{l=1}^M |J_{lk}^{SB}|^2 Q_{lk} f_l}{\sum_{l=1}^M |J_{lk}^{SB}|^2 Q_{lk}},
    \label{eq:n_k_perturb}
\end{equation}
where $Q_{lk}$ is a Lorentzian 

\begin{equation}
    Q_{lk} = \frac{\gamma}{\left(\epsilon_l^B - \epsilon_k^S\right)^2 + \gamma^2/4},
\end{equation}
and $f_l$ is the bath's thermal occupation of the eigenmode with energy $\epsilon_l^B$ at temperature $T_B$. The matrix elements $J_{lk}^{SB}$ are defined by re-writing the system-bath interaction in the energy eigenbasis

\begin{equation}
    H_{SB} = \sum_{l=1}^M\sum_{k=1}^N J_{lk}^{SB} c_l^\dagger c_k + \hc.
\end{equation}

Next, we take $\gamma\to0$. In this limit, the exchange rate with the bath simply becomes a delta function $Q_{lk}\to 2\pi\delta(\epsilon_l^B-\epsilon_k^S)$. This in turn collapses the sum in~\eqref{eq:n_k_perturb} to a single term and we recover perfect thermalization $n_k = f_k$, as long as there is an energy mode of the bath that matches each one in the system $\epsilon_l^B=\epsilon_k^S$. The size of the bath plays a crucial role here, since a larger bath is more likely to have a broad enough energy spectrum that contains all the energy levels in the system~\cite{reichental2018}. Moreover, it is usually beneficial to make the bath's Hamiltonian identical to the one describing the system, since this guarantees that their spectrums will have significant overlap. Any finite $\gamma$ broadens $Q_{lk}$ into a Lorentzian, which is more forgiving of energy mismatches, but also introduces a small deviation from thermal occupation numbers.  

\begin{figure*}[tp]
\begin{center}
\includegraphics[width = \textwidth]{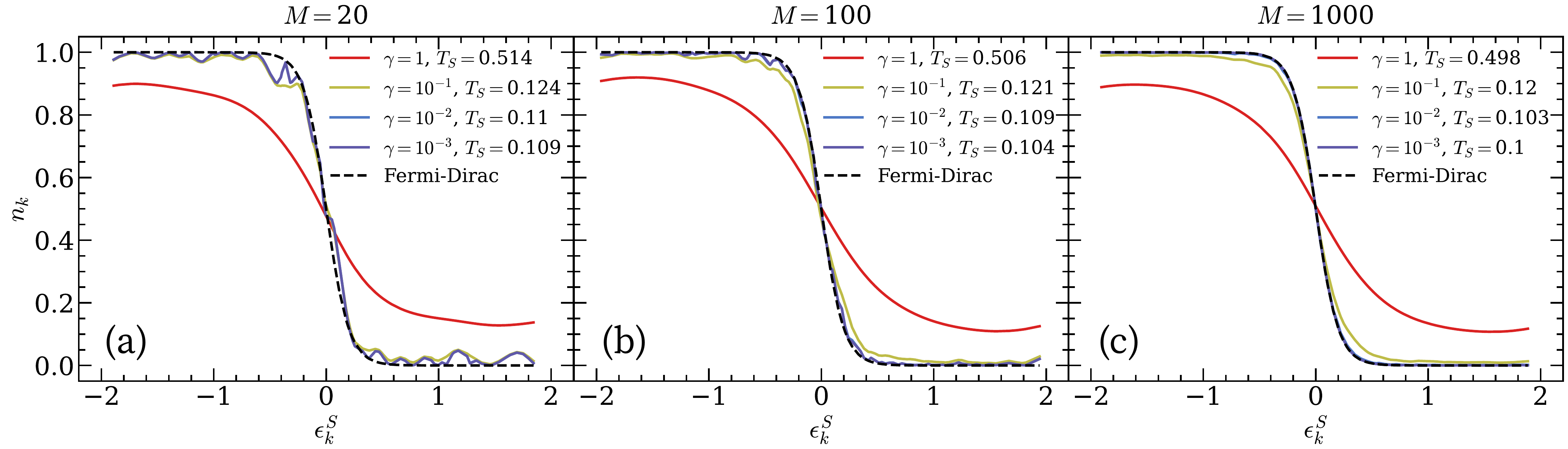}
\caption{Perturbation theory for the SYK2 system in the limit of $g\to0$ and at low temperature $T_B = 0.1$. Steady state occupation numbers $n_k$ of the single-particle energy levels $\epsilon_k^S$ match the Fermi-Dirac distribution in~\eqref{eq:fermi-dirac} if $\gamma\ll T_B$. In this limit, the fit improves substantially for larger bath sizes $M$.}
\label{fig:fig3}
\end{center}
\end{figure*}

We numerically verify our claims in the case of low temperature $T_B = 0.1$, which is the hardest setup for thermalization. Our results are presented in~\figref{fig:fig3}. First, we see that for our smallest bath, there are significant deviations in the occupation numbers even as $\gamma\to0$. Second, we notice that the increase in bath size at large values of $\gamma$ does almost nothing for thermalization. This suggests that lowering $\gamma$ should be a priority over increasing the bath size in actual simulations. In fact, having good convergence to $T_B$ requires making $\gamma\ll T_B$. The same conclusion was reached in Ref.~\cite{reichental2018}. 

Unfortunately, taking the limit of both $g$ and $\gamma$ to zero can be problematic for models that do not have closed-form analytic solutions and rely on time-evolution methods to find the steady state. We verified numerically that the Liouvillian gap scales according to~\eqref{eq:gap}. Furthermore, the average spectral gap depends on system size as $\Delta_{\mathcal{L}}\sim1/N^2$, which is better than the expected $1/N^3$ scaling for integrable models~\cite{znidaric2015}. This is likely due to the non-local all-to-all interactions of SYK2. Therefore, the convergence rate to a steady state solution would be extremely slow for large systems with small parameters $g$ and $\gamma$. However, for most practical purposes, such as quantum transport, reaching perfect thermalization at low temperatures is not essential. In fact, it is often sufficient to cool down the system to a temperature in the ballpark of $T_B$ and then use a thermometry procedure to determine the exact $T_S$~\cite{zanoci2021}. Moreover, temperatures of interest can be on the order of $J_S$, which are significantly easier to achieve with this setup.

\subsection{Interacting Models}
\label{sec:interacting}

\begin{figure*}[tp]
\begin{center}
\includegraphics[width = \textwidth]{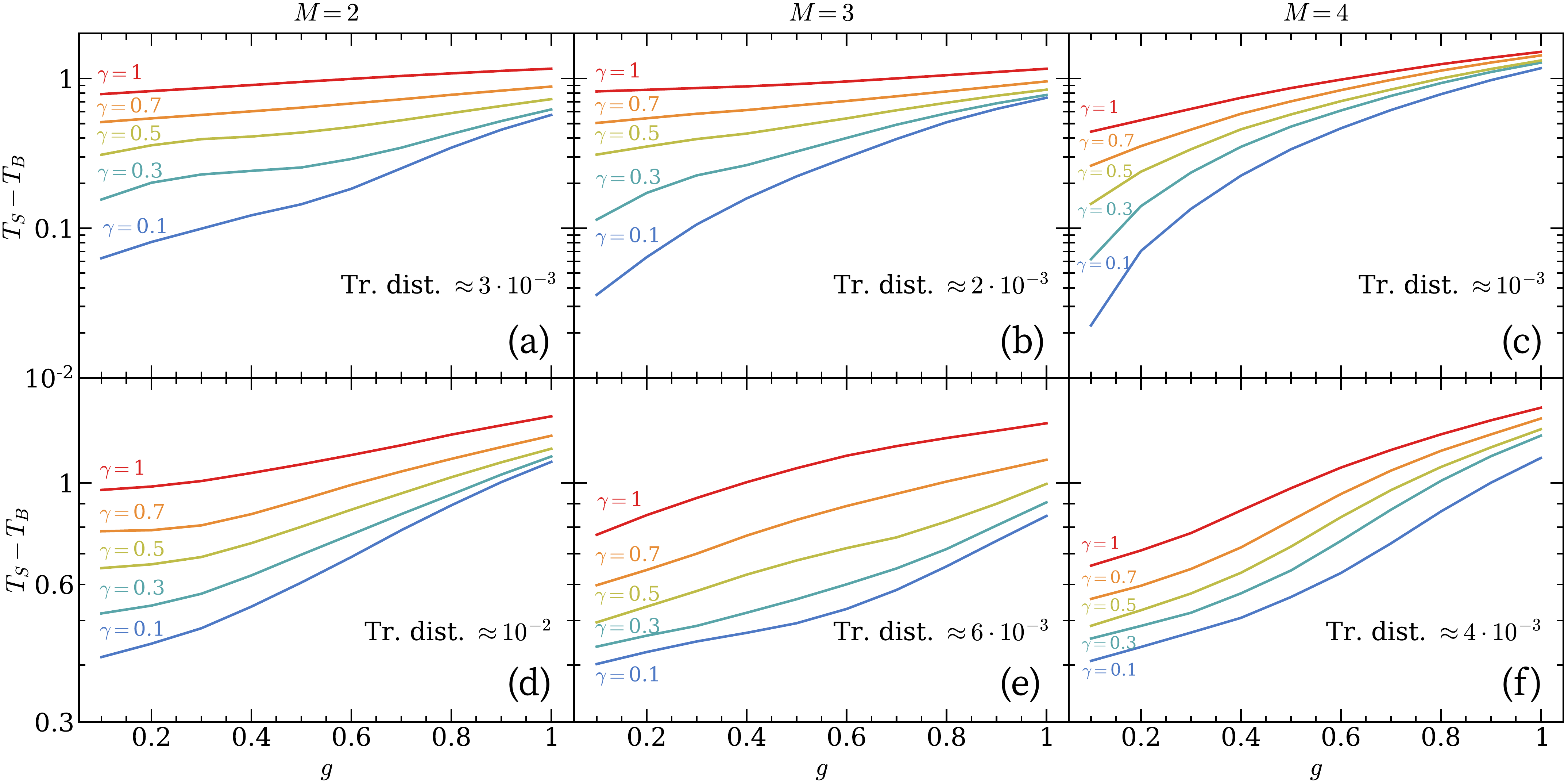}
\caption{Thermalization results for the XZ model coupled to a boundary bath at temperatures (a-c) $T_B = 1$ and (d-f) $T_B = 0.1$. For intermediate bath temperatures (top), the system approaches a temperature close to $T_B$ in the limit of weak system-bath coupling $g$ and damping $\gamma$. At low bath temperatures (bottom), the system's final temperature is limited by its gap $\Delta=0.51$. The reference trace distances correspond to the steady states with the lowest temperature.}
\label{fig:fig4}
\end{center}
\end{figure*}

\subsubsection{Gapped XZ Model}
\label{sec:XZ}

Our second model is a uniform XZ spin chain in a transverse magnetic field

\begin{align}
    H_B + H_S &= \sum_{i\neq M} \left(J_x \sigma_i^x\sigma_{i+1}^x + J_z \sigma_i^z\sigma_{i+1}^z\right) + h_x\sum_{i=1}^{M+N} \sigma_i^x,\\
    H_{SB} &= J_x \sigma_{M}^x\sigma_{M+1}^x + J_z \sigma_{M}^z\sigma_{M+1}^z,
\end{align}
where $\sigma_i^{x, z}$ denote Pauli matrices at site $i$. We choose $(J_x, J_z, h_x) = (1, 0.75, 0.21)$, which places the model in a non-integrable, quantum chaotic regime~\cite{ye2019emergent}. The model is known to have a small, but finite energy gap equal to $\Delta = 0.51$~\cite{zanoci2021}. In order to thoroughly investigate thermalization in this model, we restrict ourselves to small systems $N=20$ and baths $M=2,3,4$, such that we can reliably reach the steady state without convergence issues.

As mentioned in~\secref{sec:setup}, we simulate the time evolution of our density matrix using the TEBD algorithm~\cite{vidal2003efficient,vidal2004efficient,schollwock2011density,paeckel2019}. During the evolution, we restrict the amount of built-up entanglement by truncating the matrices to a maximum bond dimension of $\chi$. We start with a larger bond dimension $\chi=256$ during the early stages, when we have rapid entanglement growth, and then slowly decrease it to $\chi=64$ as we approach the steady state. We choose a time step of $\delta t=0.05$, which is small enough so as to not dominate over the truncation error, and evolve up to late times $t=4000$ for the smallest values of $g$ and $\gamma$. Additionally, we implement an annealing procedure which saves a lot of computational time when sweeping these parameters. We use the steady state results from larger values of $g$ and $\gamma$ as the starting point of the time evolution at lower parameter values.

We present our findings in~\figref{fig:fig4}. Qualitatively, they are similar to the non-interacting case. The temperature is uniform throughout the system, away from the boundaries. At $g=\gamma=1$, the local temperature is much larger than the driving temperature, which is consistent with previous studies~\cite{znidaric2010,vznidarivc2011transport,mendoza2015,mendoza2019asymmetry,zanoci2021}. Its value decreases continuously with $g$ and $\gamma$. For intermediate bath temperatures $T_B=1$, the system's temperature $T_S$ can be within $1\%$ of the target at the smallest $g=\gamma=0.1$. On the other hand, for low temperatures $T_B=0.1$, the final bulk temperature is strictly limited by the model's gap $T_S\approx\Delta$. Notice that this was not the case for the SYK2 model, where the gap is of order $1/N$ and still below the bath temperature. Although we only increase $M$ by a few sites, it has a substantial effect on the trace distance (see~\figref{fig:fig4}). Larger baths lead to faster and more robust convergence to the steady state. However, they are not helpful in lowering the system's temperature below $\Delta$.

\subsubsection{Gapless $\mathbb{Z}_3$ Chiral Clock Model}
\label{sec:CCM}

As our third model, we investigate a $\mathbb{Z}_3$ chiral clock model in a one-dimensional chain~\cite{samajdar2018,whitsitt2018} whose Hamiltonian can be written as
\begin{align}
    H_B + H_S &= - J\sum_{i\neq M} \sigma_i \sigma^\dag_{i+1} e^{i \theta} - f\sum_{i=1}^{M+N} \tau_i e^{i \phi} + \hc,\\
    H_{SB} &= - J \sigma_M \sigma^\dag_{M+1} e^{i \theta} + \hc,
\end{align}
where $\tau_i$ and $\sigma_i$ are the local three-state `spin' operators at site $i$ in the following matrix representations:
\begin{equation}
    \tau = 
    \begin{pmatrix}
    1 & 0 & 0\\
    0 & \omega & 0\\
    0 & 0 & \omega^2
    \end{pmatrix}, \quad
    \sigma = 
    \begin{pmatrix}
    0 & 1 & 0\\
    0 & 0 & 1\\
    1 & 0 & 0
    \end{pmatrix}, \quad
    \omega = e^{\frac{2\pi i}{3}}.
\end{equation}
These operators satisfy $\tau^3=\sigma^3=I$ and $\sigma\tau = \omega\tau\sigma$. Notice that the local Hilbert space dimension is now $d=3$, which significantly increases the computational complexity of the problem. As a result, we limit our analysis to small systems of size $N=16$ and $M=2,3$. As our Hamiltonian parameters, we choose $(J, f, \theta, \phi) = (0.5373, 0.4627, \pi/8, 0)$ such that the model is at a quantum phase transition, which has been revealed by various numerical techniques~\cite{zhuang2015phase,dai2017entanglement,samajdar2018}. Consequently, the model becomes gapless and is therefore a good target for investigating the performance of our multi-site bath scheme, with possible applications to quantum critical transport near zero temperature.

\begin{figure}[tp]
\begin{center}
\includegraphics[width = \columnwidth]{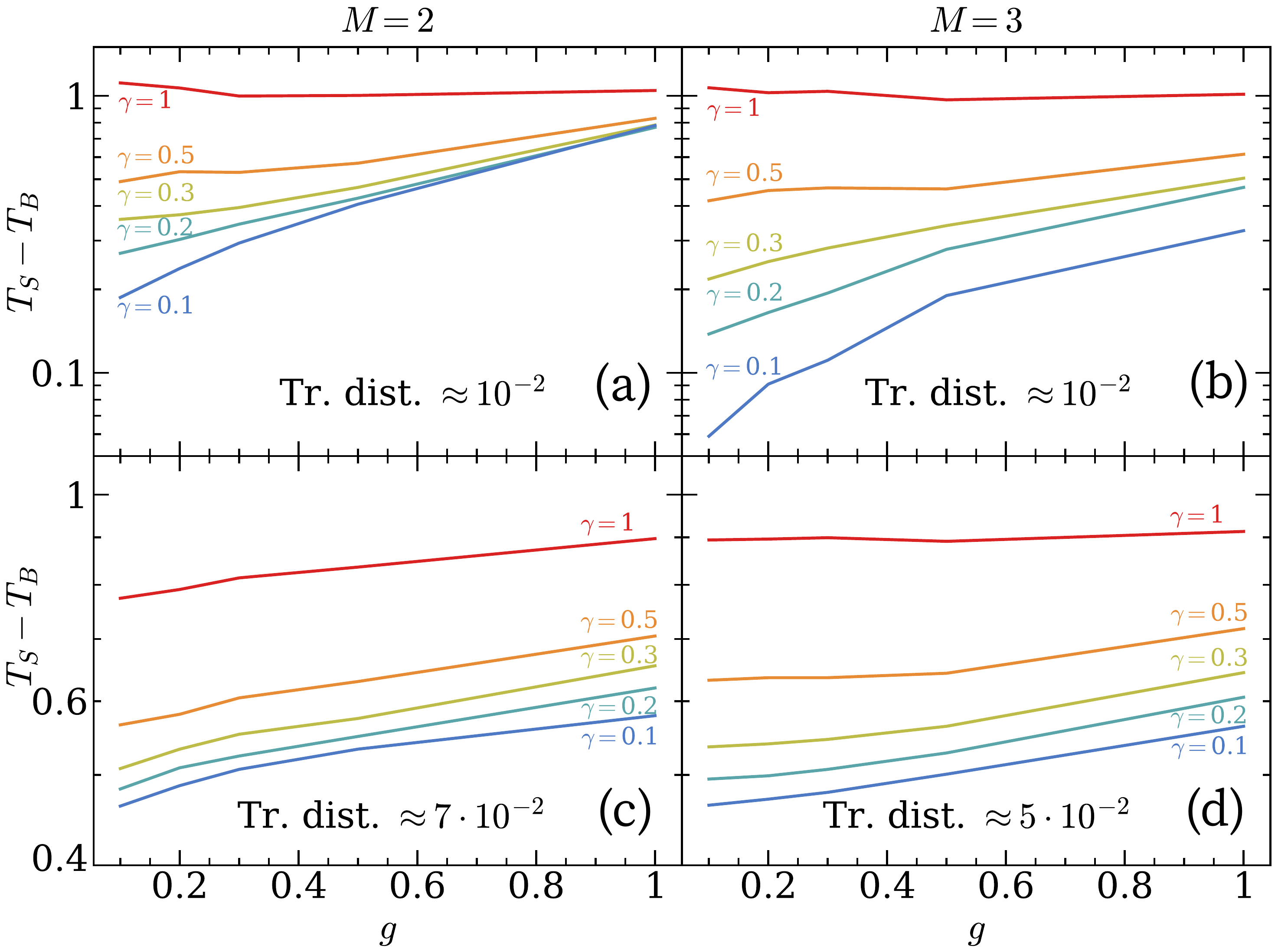}
\caption{Thermalization results for the chiral clock model coupled to a boundary bath at temperatures (a-b) $T_B = 1$ and (c-d) $T_B = 0.1$. For intermediate bath temperatures (top), the system approaches a temperature close to $T_B$ in the limit of weak system-bath coupling $g$ and damping $\gamma$. At low driving temperatures (bottom), the system's final temperature is significantly above $T_B$. The trace distances indicated in each panel are for the steady states with the lowest temperature.}
\label{fig:fig5}
\end{center}
\end{figure}

The general approach to obtaining the steady state for this model is similar to that of the XZ spin chain. We employ the same annealing procedure and set the minimal bond dimension to $\chi=81$. The convergence is noticeably slower, so we evolve the initial state up to $t=2\cdot10^4$ for the smallest values of $g$ and $\gamma$. The results for the chiral clock model are displayed in~\figref{fig:fig5}. In terms of final bulk temperature, they resemble our observations for the XZ model at intermediate temperature $T_B = 1$. However, in the low temperature $T_B = 0.1$ case, the final bulk temperature does not improve past $T_S \approx 0.5$, even though the system is gapless. Nevertheless, there are slight improvements in the trace distance when expanding the bath from $M=2$ sites to $M=3$ sites. It is quite possible that cooling a gapless model may require much larger baths and longer convergence times, beyond what is accessible numerically with the current setup. Another explanation is that there may be an emergent energy scale that prohibits the Lindblad operators from cooling the system below that scale. We elaborate more on this in~\secref{sec:discussion}.

\subsection{Application to Low-Temperature Transport}

In this section, we put to use the lessons learned about thermalization in open systems to explore some previously inaccessible physics of the XZ model. More specifically, we study its energy transport at low temperatures. The boundary-driven setup consists of a system coupled to two baths at its ends (see~\figref{fig:fig1}(b)), which imposes a temperature imbalance driving the system out of equilibrium. The left and right baths are maintained at temperatures $T_{L,R}=T_B\pm\delta T$ using the same Lindblad operators as before. The temperature offset is taken to be small $\delta T = 0.1T_B$, so that we remain in the linear-response regime. Under this assumption, we can assign a local temperature for the system weakly perturbed from equilibrium. This temperature varies slowly and linearly in the bulk, as investigated in more detail in our previous work~\cite{zanoci2021}, and hence the system's temperature $T_S$ can still be defined according to~\eqref{eq:system_temp}.

\begin{figure}
\begin{center}
\includegraphics[width=\columnwidth]{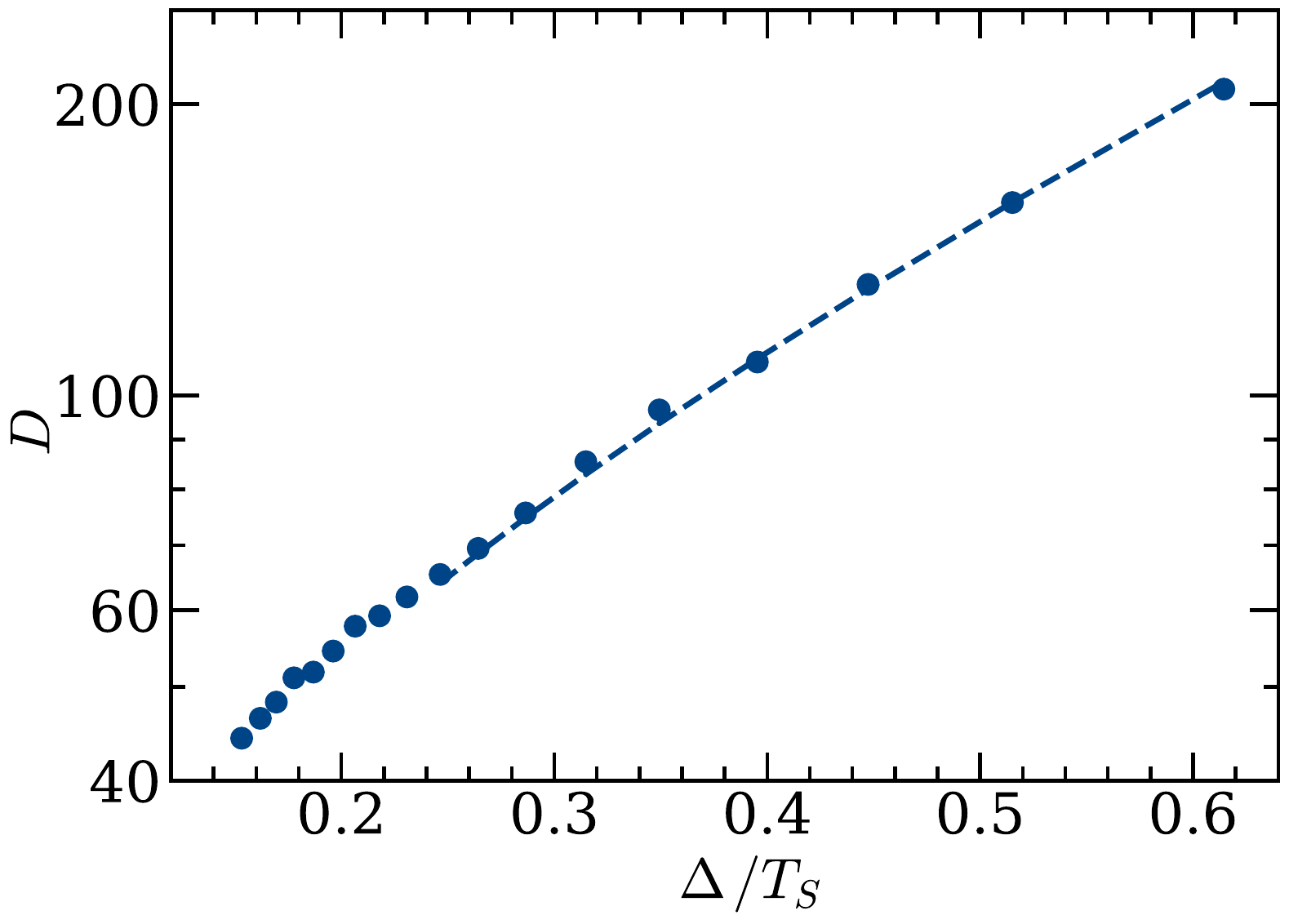}
\caption{Temperature dependence of the energy diffusion constant $D$ for the XZ model. Symbols represent numerical values obtained via our open system setup and the dashed line is fit to \eqref{eq:D_semi_class}. At low temperatures, the diffusivity grows exponentially with inverse temperature.}
\label{fig:fig6}
\end{center}
\end{figure}

The system evolves while coupled to the baths until it reaches a non-equilibrium steady state (NESS) characterized by a uniform current flowing through the chain

\begin{equation}
\begin{split}
    j &= 2J_xJ_z\langle(\sigma_{i-1}^x\sigma_i^y\sigma_{i+1}^z - \sigma_{i-1}^z\sigma_i^y\sigma_{i+1}^x)\rangle \\
    &- h_xJ_z\langle(\sigma_{i-1}^z\sigma_i^y - \sigma_i^y\sigma_{i+1}^z)\rangle.
    \label{eq:current}
\end{split}
\end{equation}
The XZ model exhibits diffusive energy transport $j=-D\nabla E$, where $\nabla E$ is the energy density gradient and $D$ is the diffusivity. The temperature dependence of this diffusion constant has been previously studied in the regime of intermediate and high temperatures~\cite{zanoci2021}. At low temperatures, the semi-classical kinetic theory predicts an exponential increase in diffusivity for gapped one-dimensional systems

\begin{equation}
    D \sim \frac{e^{2\Delta/T_S}}{\sqrt{T_S}}.
    \label{eq:D_semi_class}
\end{equation}
Unlike spin diffusion, which only relies on two-body collisions~\cite{damle1998spin,damle2005universal}, energy transport requires three-body collisions to relax the current~\cite{zanoci2021} and therefore doubles the exponent. Strictly speaking, this scaling is derived for the regime $T_S\ll\Delta$. However, since the gap for our model is so small, we find that it correctly describes the increase in diffusivity even at temperatures slightly above the gap. 

Our numerical results for a system of size $N=51$ and bath temperatures between $T_B=0.4$ and $T_B=2$ are showcased in~\figref{fig:fig6}. The diffusion constant at low temperatures matches the semi-classical prediction in~\eqref{eq:D_semi_class} remarkably well, with the only fitting parameter being the overall prefactor. In our previous studies, we managed to attain a minimum local temperature of $T_S=2.1$ in the bulk~\cite{zanoci2021}. Now we have decreased this value to $T_S=0.8$. We should point out that $g$ cannot be too small, since it would decrease the current and energy gradient to the point where they can be affected by the numerical precision of our simulations. We also emphasize that $\gamma$ should not be taken to zero, since it may cause the bulk to decouple from the boundary, resulting in a different scaling of the current~\cite{prosen2011}. Therefore, we choose $g = \gamma = 0.4$, which strikes a good balance between having well-converged NESS and reaching low temperatures.

\section{Discussion}
\label{sec:discussion}

% what we did

In this paper, we studied thermalization in open quantum systems coupled to a bath at their boundary. We investigated the emergent steady state in these systems and determined whether its associated temperature matches the driving temperature of the bath. Our analysis was based on three models: a free-fermion SYK2 cluster, a gapped XZ spin chain, and a gapless chiral clock model. For all these models, we found that the system's temperature in the default parameter regime was much higher than the target bath temperature. However, upon increasing the bath size $M$ and lowering the system-bath coupling $g$ and the bath relaxation rate $\gamma$, we saw that the two temperatures start to agree. In fact, using perturbation theory for the SYK2 model, we were able to show that the system reaches perfect thermalization at our desired temperature in the limit $g, \gamma \to 0$ and $M\gg N$. These results carry over to the interacting models as well, where we found that even baths comprised of a few sites can approximately impose the correct temperature on the system in the limit of weak coupling and damping. 

% practical implications for low-temperature transport

We demonstrated the applicability of our method by computing the low-temperature energy diffusion constant of the XZ model in an open system setup. We were able to reach temperatures much lower than in our previous work~\cite{zanoci2021} and showed that the diffusivity scales exponentially with inverse temperature, as predicted by a semi-classical calculation for gapped one-dimensional systems~\cite{damle1998spin,damle2005universal,zanoci2021}.

% minimal temperature and gap

Although our approach can successfully thermalize the system under the aforementioned conditions, there is still a minimal temperature below which the system cannot be cooled with the current setup. For the XZ model, this temperature seems to exactly match its energy gap. However, this is not true for all models. For example, we have shown that an Ising model in a mixed field, which has a relatively large gap, can be cooled far below this energy scale~\cite{zanoci2021}. At present, we do not fully understand what determines this minimal temperature, but we conjecture that it is on the order of $\min(\Delta, J)$, where $\Delta$ is the energy gap and $J$ is the typical interaction strength of the model. Moreover, it is unclear whether the limiting gap is that of the system or the bath, since both are represented by the same Hamiltonian in this case. It is difficult to distinguish between the two scenarios, because changing the bath Hamiltonian usually leads to significantly less efficient cooling at low temperatures and hence a larger minimal temperature, even if the gap of the bath Hamiltonian stays roughly the same. This is in agreement with our findings for the non-interacting model, where we showed that the energy spectra of the system and bath must closely match. Surprisingly, we also found a minimal accessible temperature for the gapless chiral clock model. A possible explanation involves the emergence of a new energy scale in the system, defined in terms of Luttinger liquid parameters~\cite{giamarchi2003quantum}, which sets its effective temperature under Lindblad dynamics. A direction of future research would be to formulate a general framework for constructing efficient baths and dynamics that could circumvent this limitation.

% mesoscopic leads 
Our study also highlights some other shortcomings of the Lindblad approach. One has to take the limit of weak coupling and damping, in addition to making the baths a substantial fraction of the system. This can become impractical for larger and more complex systems. Recently, there has been a lot of progress towards engineering mesoscopic leads with better thermalization properties. Refs.~\cite{brenes2020,lacerda2022} described tensor network algorithms for boundary-driven thermal machines, where mesoscopic baths are systematically approximated by a finite number of damped fermionic modes. Concurrently, Refs.~\cite{rams2020,wojtowicz2020} introduced a mixed spatial-energy basis for fermionic systems coupled to mesoscopic leads, which significantly lowers the required bond dimension in tensor network simulations. Tensor network methods have also been developed to implement other quantum master equations, such as the Redfield equation~\cite{xu2019}, although only for short evolution times. Given this abundance of new methods, it would be interesting to apply them to larger scale problems of interest, such as low-temperature transport, and compare them to our approach.    

\begin{acknowledgments}
We would like to thank Christopher White for valuable discussions. C.Z. acknowledges financial support from the Harvard-MIT Center for Ultracold Atoms through the NSF Grant No. PHY-1734011. Y.Y. acknowledges the U.S. Department of Energy (DOE), Office of Science, Office of Advanced Scientific Computing Research (ASCR) Quantum Computing Application Teams program, for support under fieldwork proposal number ERKJ347.
\end{acknowledgments}

\bibliographystyle{apsrev4-2}
\bibliography{references}

\appendix

\section{Multi-site Lindblad Operators}
\label{sec:appendixA} 

Our goal is to construct a super-operator $\mathcal{L}_B$ from a set of Lindblad operators $\{L_{jk}\}$ such that it drives the $M-$site bath to a Gibbs state at temperature $T_B$ and chemical potential $\mu_B$, i.e., $\mathcal{L}_B(\rho_B)=0$ where 

\begin{equation}
    \rho_B = \frac{e^{-(H_B-\mu_BN_B)/T_B}}{\Tr\left(e^{-(H_B-\mu_BN_B)/T_B}\right)},
\end{equation}
and $N_B$ is the total spin or particle number operator of the bath. We therefore require that $\rho_B$ is a unique eigenvector of $\mathcal{L}_B$ with eigenvalue $0$. However, this condition does not fully fix the jump operators, as it only ensures that the steady state is correct. One can additionally require that all the other modes decay at the same rate~\cite{prosen2009matrix}, which results in the fastest convergence to $\rho_B$. Alternatively, one could impose the detailed-balance condition between the energy levels of $H_B$, which may lead to better thermalization in certain regimes~\cite{palmero2019}. For our models, we find that both approaches work equally well even at low temperatures. 

In the case of free fermions, the number of Lindblad operators required to thermalize the bath scales linearly with its size, while for a generic spin system, this number scales exponentially with $M$. This may seem problematic at first, since it would severely restrict the size of the bath that can be implemented in practice. However, as we show in the main text, even a relatively small bath can result in good thermalization under the right conditions. The real bottleneck is in designing a compact tensor network representation of the Liouvillian $\mathcal{L_B}$, which can be efficiently applied to the bath without generating too much entanglement during time evolution. A potential avenue of research would be to leverage the Product Spectrum Ansatz~\cite{martyn2019,sewell2022} to design dissipators that only approximately thermalize larger baths.

\subsection{Non-interacting Hamiltonian}

If the bath Hamiltonian is quadratic in the fermion creation and annihilation operators, the jump operators can be made linear~\cite{mahajan2016entanglement,zanoci2016entanglement,reichental2018}. We first write the Hamiltonian (\eqref{eq:SYK_ham}) in the energy eigenbasis by diagonalizing the interaction matrix $J^B = V^\dagger \epsilon^B V$, where $\epsilon^B=\text{diag}(\epsilon_1^B, \epsilon_2^B, \ldots, \epsilon_M^B)$ and $V$ is unitary. We recover~\eqref{eq:SYK_ham_energy} with $c_k = \sum_{j=1}^M V_{kj}c_j$. Note that the new operators also satisfy the canonical anti-commutation relations $\{c_k^\dagger, c_{l}\}=\delta_{kl}$. The thermal density matrix can be written in this basis as well

\begin{equation}
    \rho_B = \prod_{k=1}^M \left(f_kc_k^\dagger c_k + (1-f_k)c_kc_k^\dagger\right),
\end{equation}
where $f_k$ are the equilibrium occupation numbers defined in~\eqref{eq:fermi-dirac}. For each mode $k$, we introduce two jump operators that either add a fermion at a rate $\gamma f_k$ or remove a fermion at a rate $\gamma (1-f_k)$

\begin{align}
    L_{\text{in},k} &= \sqrt{\gamma f_k} c_k^\dagger,\\
    L_{\text{out},k} &= \sqrt{\gamma (1-f_k)} c_k.
\end{align}
The rates are chosen to satisfy the detailed-balance condition $f_k/(1-f_k) = e^{-(\epsilon_k^B-\mu_B)/T_B}$. We can further check that all the terms in the Lindblad equation exactly cancel

\begin{equation}
\begin{split}
    &L_{\text{in},k}\rho_B L_{\text{in},k}^\dagger = L_{\text{out},k}^\dagger L_{\text{out},k} \rho_B = \rho_B L_{\text{out},k}^\dagger L_{\text{out},k} \\
    &= \gamma f_k(1-f_k) c_k^\dagger c_k \prod_{l\neq k}
    \left(f_{l}c_{l}^\dagger c_{l} + (1-f_{l})c_{l}c_{l}^\dagger\right),
\end{split}
\end{equation}
\begin{equation}
\begin{split}
    &L_{\text{out},k}\rho_B L_{\text{out},k}^\dagger = L_{\text{in},k}^\dagger L_{\text{in},k} \rho_B = \rho_B L_{\text{in},k}^\dagger L_{\text{in},k} \\
    &= \gamma f_k(1-f_k) c_k c_k^\dagger \prod_{l\neq k}
    \left(f_lc_l^\dagger c_l + (1-f_l)c_lc_l^\dagger\right).
\end{split}
\end{equation}
Thus we conclude that $\rho_B$ is indeed a fixed point under the dynamics generated by these jump operators. 

\subsection{Interacting Hamiltonian}

For interacting spin systems, we extend the two-site Lindblad operators construction in Refs.~\cite{prosen2009matrix,vznidarivc2011transport,mendoza2015,mendoza2019asymmetry,zanoci2021} to systems of arbitrary size. Consider the general case where each of the $M$ spins has a local Hilbert space dimension $d$. We begin by diagonalizing the density matrix $\rho_B = V^\dagger W V$, where $W=\text{diag}(W_0, W_1, \ldots, W_{d^M-1})$ and $V$ is unitary. Define a set of $d^{2M}$ operators $\tilde{L}_{jk} \in \mathbb{R}^{d^M\times d^M}$

\begin{equation}
    \tilde{L}_{jk} = \sqrt{\gamma W_j} E_{jk}, \quad 0\leq j,k < d^M,
\end{equation}
where $E_{ab}$ is the matrix unit with a $1$ in row $a$ and column $b$ as its only non-zero entry. Here $\gamma$ quantifies the overall strength of the bath damping. It is easy to verify that 

\begin{align}
    \tilde{L}_{jk}W\tilde{L}_{jk}^\dagger &= \gamma W_j W_k E_{jj},\\
    \tilde{L}_{jk}^\dagger\tilde{L}_{jk}W &= W\tilde{L}_{jk}^\dagger\tilde{L}_{jk} = \gamma W_j W_k E_{kk}.
\end{align}
Therefore we have 

\begin{equation}
\begin{split}
    \tilde{\mathcal{L}}_B(W) &= \sum_{j,k=0}^{d^M-1} \Big(\tilde{L}_{jk}W \tilde{L}_{jk}^\dagger - \frac{1}{2}\tilde{L}_{jk}^\dagger \tilde{L}_{jk}W - \frac{1}{2}W\tilde{L}_{jk}^\dagger \tilde{L}_{jk}\Big)\\
    &= \gamma\sum_{j,k=0}^{d^M-1} W_jW_k\left(E_{jj} - E_{kk}\right) = 0,
\end{split}
\end{equation}
and we can multiply this expression by $V^\dagger$ and $V$ on the left and right sides, and use the identity $VV^\dagger = I$ to deduce that

\begin{equation}
    \sum_{j,k=0}^{d^M-1} \Big(L_{jk}\rho_B L_{jk}^\dagger - \frac{1}{2}L_{jk}^\dagger L_{jk}\rho_B - \frac{1}{2}\rho_BL_{jk}^\dagger L_{jk}\Big) = 0,
\end{equation}
with $L_{jk} = V^\dagger \tilde{L}_{jk}V$. Hence we can use these new Lindblad operators $L_{jk}$ to construct a super-operator satisfying $\mathcal{L}_B(\rho_B)=0$. Moreover, since $V$ is unitary, the eigenvalues of $\mathcal{L}_B$ will be the same as those of $\tilde{\mathcal{L}}_B$. We can compute the latter using the vectorized representation of the Liouvillian~\cite{zwolak2004,weimer2019simulation,landi2022}

\begin{equation}
    \tilde{\mathcal{L}}_B = \sum_{j,k=0}^{d^M-1} \Big(\tilde{L}_{jk}^*\otimes \tilde{L}_{jk} - \frac{1}{2}I\otimes \tilde{L}_{jk}^\dagger \tilde{L}_{jk} - \frac{1}{2}\tilde{L}_{jk}^T\tilde{L}_{jk}^*\otimes I\Big),
\end{equation}
where

\begin{align}
    \tilde{L}_{jk}^*\otimes \tilde{L}_{jk} &= \gamma W_j E_{j(d^M+1),k(d^M+1)},\\
    I\otimes \tilde{L}_{jk}^\dagger \tilde{L}_{jk} &= \gamma W_j \sum_{i=0}^{d^M-1} E_{k+i\cdot d^M,k+i\cdot d^M},\\
    \tilde{L}_{jk}^T\tilde{L}_{jk}^*\otimes I &= \gamma W_j \sum_{i=0}^{d^M-1} E_{i+k\cdot d^M,i+k\cdot d^M},
\end{align}
and $I$ denotes the $d^M\times d^M$ identity matrix. It is straightforward to check that $\tilde{\mathcal{L}}_B$ has exactly one zero eigenvalue and the remaining eigenvalues are equal to $-\gamma$, since $\sum_{j=0}^{d^M-1}W_j = \Tr(\rho_B) = 1$. Hence our construction leads to the fastest relaxation to the target density matrix~\cite{prosen2009matrix}. 

A different set of dissipators, satisfying detailed-balance relations, has also been proposed~\cite{palmero2019}. Using our notation above, these operators take the form

\begin{equation}
    \tilde{L}_{jk} = \sqrt{\frac{\gamma W_j}{W_j+W_k}} E_{jk}, \quad 0\leq j\neq k < d^M.
\end{equation}
We can again verify that they drive the bath to its correct thermal state

\begin{equation}
    \tilde{\mathcal{L}}_B(W) = \gamma\sum_{j,k=0}^{d^M-1} \frac{W_j W_k}{W_j+W_k}\left(E_{jj} - E_{kk}\right) = 0,
\end{equation}
since 

\begin{align}
    \tilde{L}_{jk}W\tilde{L}_{jk}^\dagger &= \gamma \frac{W_j W_k}{W_j+W_k} E_{jj},\\
    \tilde{L}_{jk}^\dagger\tilde{L}_{jk}W &= W\tilde{L}_{jk}^\dagger\tilde{L}_{jk} = \gamma \frac{W_j W_k}{W_j+W_k} E_{kk}.
\end{align}
The final jump operators are given by $L_{jk} = V^\dagger \tilde{L}_{jk}V$.

\end{document}